# The electronic structure of CeCoIn$_5$ from angle-resolved photoemission spectroscopy

A. Koitzsch[1], I. Opahle[1], S. Elgazzar[1], S. V. Borisenko[1], J. Geck[1], V. B. Zabolotnyy[1], D. Inosov[1], H. Shiozawa[1], M. Richter[1], M. Knupfer[1], J. Fink[1,3], B. Büchner[1], E. D. Bauer[2], J. L. Sarrao[2], and R. Follath[3]

[1] *IFW Dresden, P.O.Box 270116, D-01171Dresden, Germany*

[2] *Los Alamos National Laboratory, Los Alamos, New Mexico 87545, USA*

[3] *BESSY GmbH, Albert-Einstein-Strasse 15, 12489 Berlin, Germany*

**Abstract**: We have investigated the low-energy electronic structure of the heavy fermion superconductor CeCoIn$_5$ by angle-resolved photoemission and band structure calculations. We measured the Fermi surface and energy distribution maps along the high-symmetry directions at hν = 100 eV and T = 25 K. The compound has quasi two-dimensional Fermi surface sheets centered at the M-A line of the Brillouin zone. The band structure calculations have been carried out within the local density approximation where the 4f electrons have been treated either localized or itinerant. We discuss the comparison to the experimental data and the implications for the nature of the 4f electrons at the given temperature.

The heavy-fermion superconductor CeCoIn$_5$ [1] has been the focus of intense research due to its unusual properties such as field-induced quantum criticality [2], unconventional superconductivity [3], and a phase transition within the superconducting state associated with either a Fulde-Ferrell-Larkin-Ovchinikov state or magnetic order [4, 5, 6].

Below ca. T* = 45 K a coherent heavy mass quasiparticle band starts to form [7], which decreases the magnetic scattering rate of the charge carriers and leads to an increase of the electrical conductivity. The effective mass increases significantly below T* . It reaches momentum integrated enhancement factors of 20 at 10 K [8] and 50 at temperatures << 1 K in high magnetic fields [9] for particular Fermi surface branches as determined by optical spectroscopy and de Haas van Alphen experiments, respectively. In the zero temperature limit, phase transitions can be achieved by tuning the external parameters pressure [10] and magnetic field [2, 11] which indicates quantum critical behavior. The non Fermi-liquid behavior accompanying the quantum critical region of the phase diagram shows signatures of antiferromagnetic fluctuations which might also be involved in the mechanism of the pairing.

Another reason of the importance of CeCoIn$_5$ and its relative compounds CeIrIn$_5$ and CeRhIn$_5$ rests on a remarkable analogy: the crystal structure of all of these materials is quasi-two-dimensional being comprised of CeIn$_3$ and $T$In$_2$ ($T$ = Co, Ir, Rh) layers. Within the series Co, Ir, Rh the anisotropy of the Co compound is strongest [9, 12] establishing a phenomenological relation between superconductivity and reduced dimensionality, and, by analogy, a relation to the cuprate superconductor problem. The cuprates are quasi two-dimensional as well and susceptible for antiferromagnetic fluctuations.

Of crucial importance for an understanding of the properties of CeCoIn$_5$ – and, generally, any other material - , is the knowledge of the electronic structure, i.e. its Fermi surface (FS), and the band structure along the crystallographic high-symmetry directions. In particular this information forms the basis of further exploration of the participation of the f-electron states to the low-energy physics and possible multiband scenarios for the superconducting order pa-rameter [13, 14, 15]. From the theoretical side, band structure calculations within the local density approximation (LDA) have proven to be a valuable tool to investigate the electronic structure of f-electron systems. At first glance this may seem surprising, because of the correlated nature of the Ce 4f electrons. However, it has been shown that the correlation effects can be accounted for by an adjustment of the scattering phase shifts of the f-electrons, which often makes little changes to the original LDA FS [16]. The LDA calculations treat the f-electrons as itinerant or localized. Well above the Kondotemperature the f electrons behave fully localized and are excluded from the Fermi surface while they contribute well below. By comparing experimental data to such calculations it is possible to judge upon the degree of localization of the f-electrons at a given temperature [17, 18]. For $CeCoIn_5$ de-Haas-van-Alphen (dHvA) experiments yielded good agreement to band structure calculations treating the 4f electrons as itinerant (4f itin) [19]. The degree of f-localization/ itinerancy is central for all 115 systems [20]. However, dHvA gives no direct information on the shape of the FS and no information on the band structure away from the Fermi energy. Furthermore it is restricted to low temperatures. In the case of $CeCoIn_5$ measurements have been done below $T = 1$ K, which is much lower than T* and deep inside the Kondo screening regime.

The Fermi surface of $CeCoIn_5$ as revealed by band structure calculations and dHvA experiments shows a nearly cylindrical and, therefore, quasi two-dimensional sheet centered at the M-A line and a number of three-dimensional complex objects [9, 12]. Similar results are found for $CeRhIn_5$ and $CeIrIn_5$ [21, 22]. The three dimensional parts of the electronic structure and also the deviations of the two dimensional sheet from a perfect cylinder complicate the analysis of the ARPES experiments considerably, since in general only the electron wave vector parallel to the sample surface is preserved during the photoemission process.

A previous ARPES investigation of the related compound $CeRhIn_5$ suggested that the 4f electrons participate in the band formation at $T = 15$ K [23]. On the other hand subsequent measurements by Fujimori et al. concluded in favor of a localized behavior of the 4f electrons in CeRhIn$_5$ and CeIrIn$_5$ [24]. Nevertheless, even if the localized behavior dominates a small itinerant component is present and may be of decisive importance for the low energy excitations [25]. An ARPES study of the two-layer compounds Ce$_2$CoIn$_8$ and Ce$_2$RhIn$_8$ found similar generic behavior of the electronic structure with respect to the one layer compounds: a cylindrical quasi two-dimensional FS branch is found centered at the M-A line and the 4f electrons show rather localized behavior [26]. In a previous publication some of us investigated the hybridization effects in CeCoIn$_5$ by ARPES [27]. From the characterisitcs of the low energy dispersion, the peakwidth and from resonant angle-resolved spectroscopy we found signatures of the conduction electrons hybridizing with the f-states at low temperatures (T = 20 K).

Here we report a combined angle-resolved photoemission (ARPES) and band structure calculation study. We present a FS map and images of the band structure along the high symmetry directions. We compare the experimental results to the LDA calculations and discuss the nature of the f-electrons.

The ARPES experiments were carried out using radiation from the U125/1-PGM beam line and an angle multiplexing photoemission spectrometer (SCIENTA SES 100) at the BESSY Synchrotron Radiation Facility. We present photoemission data taken with hν = 87 eV and hν = 100 eV with an energy resolution of 40 meV for the data presented in Fig. 5 and 70 meV for the rest and an angular resolution of 0.2°. Single crystals of CeCoIn$_5$ were grown in In flux. For the measurements the crystals were cleaved *in situ* at room temperature. The band structure calculation was performed using the scalar-relativistic version of the full potential local orbital minimum basis band-structure method [28, 29]. Technical details of the calculations have been described in Ref. [19].

Figure 1 shows a Fermi surface map at T = 25 K along with a schematics of the Brillouin zone. The cleavage surface is parallel to the Γ-M-X plane. This Fermi surface map consists of a nearly square shaped structure centered at the corner of the Brillouin zone and a more complicated "flower structure" with four leaves around the Z-point. The notation "Z" instead of "Γ" will be justified below.

The Fermi surface map presented in Fig. 1 is obviously a two-dimensional object. How is it related to the overall three dimensional FS expected for CeCoIn$_5$? At a given photon energy a FS map like the one in Fig. 1 corresponds to a hemispherical cut through the three dimensional BZ, provided that the final states are well described by a free electron parabola and the $k_z$ broadening is negligible [30]. For hν = 100 eV the radius of the hemisphere is much larger then the dimension of the BZ, therefore it is justified to consider the cut flat over the first BZ and parallel to the surface (parallel to the Γ-M-X plane). However, for higher BZ one must keep in mind, that the $k_z$ value changes compared to the first BZ. A priori it is unknown which photon energy corresponds to which $k_z$ cut. The rigorous way to examine the electronic structure is to vary the photon energy and to scan thereby the full Brillouin zone in $k_z$ direction. The absolute value of $k_z$ can then be determined by exploiting the $k_z$ periodicity. However, here we follow a different strategy. We compare the data for hν = 100 eV and in one case for hν = 87 eV to band structure calculations for two special $k_z$ values, namely the Γ- M -X plane ("Γ- plane") and the Z-A-R plane ("Z-plane"). In the left upper corner of Fig. 1 the FS map is compared to LDA calculations for the Γ-plane (blue) and the Z-plane (green) where the f-electrons are treated as localized. The FS cut for the Z-plane consists of a rounded square and a circle centered at the BZ corner and a small diamond in the zone center. On the other hand the FS for the Γ- plane is composed of a rounded diamond, a broad cross, and a large square all centered at the BZ corner. The experimental FS compares better to the Z-plane cut. In particular the inner FS sheets around the corner of the BZ have the shape of a rounded square and not of a diamond as it should be for the Γ point. However, the green circle around the BZ cor-ner is not visible in the data. The agreement of the calculation to the "flower" structure is much worse. Experimentally the intensity of this feature varies strongly from cleavage to cleavage. Furthermore we found that it stays constant with varying photon energies. Therefore, we assign it to a surface state. The small diamond structure in the zone center for the green points is not discernible in the first BZ due to the surface state but seems to be reproduced around $Z_2$. A weak structure, called "line" in the figure, is visible in the second BZ. In fact this feature is also present in the first BZ as can be seen in the EDMs (Energy Distribution Map) below. It may partly fit to the outer, squarish sheet of the $\Gamma$ – plane FS cut. Its origin is discussed below. The comparison of the experimental FS to 4f itin calculations is worse (not shown) as they require additional FS-sheets which are not observed at this point. Closer inspection of the "square" reveals that it consists of two separate bands (see Fig. 5), which we call, following an earlier notation scheme [9], α and β. The spectral weight of α and β varies along the FS because it depends on matrix element effects. These observations are consistent with previous dHvA measurements [9, 12].

Figures 2-4 compare the experimental band structure along the high-symmetry directions with theoretical results where the f-electrons are treated as localized (4f loc) or itinerant respectively. At a photon energy of hν = 100 eV the photoemission cross section favors emission from Co 3d [31]. Therefore, we compare in Figure 2 the experimental data along ZA (ΓM) (panel a) to the Co 3d orbital projected band structure (panels b-e). The weight of the Co 3d orbital is proportional to the size of the blue circles. We stress that the character of the f electrons (localized versus hybridized) has a substantial influence on the overall shape of the band structure. Therefore it is possible to derive conclusions about the nature of the f-electrons by comparing the experimental data to calculations where the f-states are treated localized or itinerant, even though the observable f-weight at hν = 100 eV is low.

In Fig. 2a the EDM along the diagonal of the Brillouin zone is presented. There is an intense band crossing $E_F$ approximately at half the distance between Z(Γ) – A(M) consistent with the Fermi surface map in Fig. 1. This crossing consists of the bright "flower" related structure and a weaker additional band. This band has a parabola like shape and is responsible for the closed square structure of the FS centered at the M-A line. Now we compare the experimental data in Fig 2a to LDA calculations for different $k_z$ cuts of the three dimensional BZ to determine which $k_z$ plane is probed with a photon energy of hν = 100 eV. Fig 2b and c show calculations for the Z-plane ($k_z = \pi/c$), Fig. 2d for the Γ plane ($k_z = 0$) and additionally Fig 2e shows an intermediate cut close to the Z-plane ($k_z = 0.85\ \pi/c$). The best agreement is observed between Fig 2a and Fig. 2b, i.e. to the Z-plane cut. The agreement to the Γ plane (Fig. 2d) is much worse. Even the comparison for the plane 15% shifted away from the Z-plane towards Γ (Fig 2e) is significantly worse than to the Z-plane itself. We conclude that with a photon energy of hν = 100 eV mainly the vicinity of the Z-plane is probed. The LDA calculation for the Z-plane is depicted in two alternatives. In Fig 2b the f-electrons have been considered as completely localized and in Fig 2c as completely itinerant. The agreement between (a) and (c) suffers from the absence of the shallow bands around Z with minima of ca. 30 meV and 150 meV in the experimental data. These states have also significant f-weight in the itinerant calculation. It follows that the localized character of the f-electrons is still pronounced at T = 25 K.

Figure 3 presents the Z-R (Γ–X) direction in an equivalent manner. Again the experimental band structure in Fig. 3a clearly resembles more the Z-plane cut (Figs. 3b, c) than the Γ-plane cut (Fig. 3d). It appears that the experimental data are shifted towards lower binding energy with respect to the calculation. The itinerant version of the Z-plane calculation requires intense flat bands around R and Z which are not observed. This is consistent with Fig. 2. Broad spectral intensity with no analogue in the theory originates from the "flower" structure.

Figure 4 shows the comparison for the A-R (M-X) direction. The main feature in the data is again the parabola shaped band centered at the BZ corner. The calculation predicts a prominent splitting of this band for the Z-plane into α and β sheets. In the bare data this splitting is hardly recognizable. By taking the second derivative of the data some splitting can be observed (not shown) but it is smaller than expected for the Z-plane. Moreover another weak band is observed (red arrow) which corresponds to the "line" feature in the FS map of Fig. 1. In this situation the comparison of the data to the Γ-plane (MX) seems promising. However, in this case the splitting should occur at higher binding energies (E = 0.5 eV) which is also not observed. Flat bands crossing the Fermi energy from above with substantial f-weight for the itinerant AR version are again not present. Additionally we present in panel e) an EDM taken with hv = 87 eV. The $k_z$ distance between the hv = 100 eV measurements and the hv = 87 eV measurement is approximately $\Delta k_z = \frac{1}{\hbar}\left(\sqrt{2m(E_{kin1}+V_0)} - \sqrt{2m(E_{kin2}+V_0)}\right)$, where $E_{kin}$ is the kinetic energy of the photoelectrons and $V_0$ is the inner potential. The inner potential is a priori unknown. For the sake of this rough estimate we adopt a value of $V_0$ = 12 eV, a number that was found for CeRu$_2$Si$_2$ [32] and is often used for HF compounds. We obtain $\Delta k_z = 0.32$ Å$^{-1}$ which needs to be compared to $\pi/c = 0.42$ Å$^{-1}$. If the 100 eV measurement probes the vicinity of the Z-plane the 87 eV measurement is significantly shifted towards Γ. Although the quality of the 87 eV data is inferior to the 100 eV data, one clearly sees that the outer parabola, the "line" feature, has become stronger. The EDM resembles now the 4f loc data in the M-X direction, although the features are shifted in energy by appr. ~ 200 meV. At this point it is important to note that there is always an inherent $k_z$ integration even for a sharp and fixed photon energy. This constrain is imposed by the finite electron escape depth. Therefore, it can happen that k-regions, which are formally somewhat away from the main $k_z$ locus of a given photon energy, mix in the photoemission yield if there are features with high enough density of states. This is a possible scenario for the appearance of the "line" feature in the 100 eV data.

To resolve the bands α and β explicitly, Fig. 5 presents high resolution EDMs parallel to A-R (M-X) corresponding to the bars labeled from 1 to 6 in Fig. 1. The high symmetry direction corresponds to panel 1. We find again the parabola shaped band at the BZ boundary (labeled α) and the "line feature" band (labeled LF). Moving to panel 2 a faint additional feature appears (labeled β) close to α. This feature becomes slightly more intense in panels 3 and 4. In panels 5 and 6 the background increases and all features become broad, making them hard to distinguish. This broadening is due to the shape of the Fermi surface: the k-space intervals corresponding to panels 5 and 6 do not cut the FS anymore at an angle of 90° therefore the width of the features must increase. The Fermi vector $k_F$ for the α band moves first slowly to higher momenta in going from panel 1 to 6 and then more rapidly (panel 5, 6) as is expected from the shape of the FS of a rounded square. Assuming that β collapses with α in panel 1 it is clear that the splitting between those bands increases in going away from the high symmetry direction. However, for the Z-plane LDA calculation the opposite behavior is expected (see Fig. 1). Exactly along A-R the splitting between the α and β-sheet is at maximum. The "line" band shows little $k_F$ variation in the k interval under scrutiny here, as is already expected from the FS map in Fig. 1. From the $k_F$ positions of the α sheet extracted from data such as in Fig.5 we determine the FS volume $S_F$ and compare it to dHvA frequencies F. We obtain a value of $S_F$ for the α sheet equivalent to F = 4500 T (+-400 T). This is to be compared with literature values of F = 4240 T [9] and F = 4566 T [12] for the orbits assigned to the Z-plane and F = 4530 T [9] and F = 5161 T [12] for the Γ-plane. There is a third extremal α orbit in between Γ and Z with F = 5401 T [9] and F = 5560 T [12]. Our value of F = 4500 T is consistent with the bulk sensitive dHvA results. But given the error bars of our measurement and the scattering of the dHvA results it is difficult to draw further conclusions from it. We abstain from extracting F for the β sheet in the same way at this point due sizable experimental uncertainty of the $k_F$ determination, but it is clear, that the FS volume of this sheet is significantly larger than that of the α sheet in qualitative agreement to the dHvA results.

The overall comparison of the experimental results and the band structure calculations shows a row of agreeing and some disagreeing aspects. The EDM along the zone diagonal compares well with the theoretical expectations for the Z-plane of the BZ (Fig. 2). The same holds true for the Z-R direction (Fig. 3). For the AR direction the agreement for the 100 eV data is less convincing due to the small α/β splitting but additional support comes from the 87 eV data which resemble the Γ plane bands. The shape of the α sheet on the FS is also in accord with theory. We find differences between experiment and theory regarding the fine structure of the β sheet. However it is fair to say that the gross features predicted by the band structure calculation are indeed found by experiment.

Generally the agreement of our data to the calculations where the f-electrons have been treated as being localized is better than to the itinerant version. This is the same result as for $CeRhIn_5$ and $CeIrIn_5$ [24]. It implies strong localized character of the f-electrons (i.e. weak hybridization) but it does not exclude the existence of a f-derived quasiparticle band. In fact, starting heavy fermion behavior is observed below T* < 45 K, e.g. from transport [1] and optical measurements [8]. Consequently, a small but finite fraction of the 4f electrons must hybridze with the conduction band electrons. A detailled study of the hybridization effects using also resonance photoemission experiments is presented in [27]. At T = 25 K the measurements are performed in an intermediate regime: it is below T* but above $T_K$, the single ion Kondo temperature. Within the two fluid model [33] the fraction of the coherently screened f-electrons increases steadily when $T_K$ is approched from above. Therefore, the conclusion of a predominantly localized behavior at 25 K is not in contradiction with our previous work [19] where an itinerant f-character was claimed at very low tempertures based on a comparison to dHvA.

ARPES is a surface sensitive technique. It might be possible that the localized behavior of the electronic structure is enhanced by the surface due to the reduced coordination numbers. To check this possibility for $CeCoIn_5$ experiments with very high or very low photon energies are in order. When discussing the experimental band structure we must also consider the possibility that certain features are suppressed by matrix element effects, or escape the attention of the observer because they are intrinsically very weak. The photoemission matrix element depends sensitively on the photon energy and on the polarization conditions. However, we did not find indications of additional bands for other photon energies and different polarization geometries. Sometimes spurious intensity is found inside the "square".

In summary we investigated the electronic structure of $CeCoIn_5$ by ARPES and by LDA based band structure calculations. The main feature of the electronic structure of the compound is the quasi-cylindrical FS sheet centered at the M-A line of the BZ. It originates from an electron like, parabola shaped band with a minimum around E = -0.6 eV. The comparison of the FS map and EDMs along the high symmetry directions at photon energies of hν = 100 eV and hν = 87 eV to the band structure calculations allows a consistent assignment of the bands. At hν = 100 eV a $k_z$ cut in the vicinity of the Z-A-R plane is probed, whereas for hν = 87 eV it is a cut closer to the Γ-M-X plane. The FS volume of the α sheet determined from the map is, within error bars, in agreement with dHvA experiments. No indications of itinerant f bands are found. It is concluded that the 4f electrons behave under the given experimental conditions predominantly localized.

We acknowledge helpful discussions with S. L. Molodtsov, J. D. Denlinger and J. W. Allen and technical support by R. Hübel, S. Leger, and R. Schönfelder.
The work was supported by the DFG via SFB 463.

# References

[1] C. Petrovic, P. G. Pagliuso, M. F. Hundley, R. Movshovich, J. L. Sarrao, J. D. Thompson, Z. Fisk, and P. Monthoux, J. Phys.: Condens. Matter **13**, L337 (2001)

[2] A. Bianchi, R. Movshovich, I. Vekhter, P. G. Pagliuso, and J. L. Sarrao, Phys. Rev. Lett. **91**, 257001 (2003)

[3] R. Movshovich, M. Jaime, J. D. Thompson, C. Petrovic, Z. Fisk, P. G. Pagliuso, and J. L. Sarrao , Phys. Rev. Lett. **86**, 5152 (2001)

[4] A. Bianchi, R. Movshovich, N. Oeschler, P. Gegenwart, F. Steglich, J. D. Thompson, P. G. Pagliuso, and J. L. Sarrao, Phys. Rev. Lett. **89** 137002 (2002)

[5] B.-L. Young, R. R. Urbano, N. J. Curro, J. D. Thompson, J. L. Sarrao, A. B. Vorontsov, and M. J. Graf, Phys. Rev. Lett. **98** 036402 (2007)

[6] V. F. Mitrovic, M. Horvati, C. Berthier, G. Knebel, G. Lapertot, and J. Flouquet, Phys. Rev. Lett. **97**, 117002 (2006)

[7] S. Nakatsuji, S. Yeo, L. Balicas, Z. Fisk, P. Schlottmann, P. G. Pagliuso, N. O. Moreno, J. L. Sarrao, and J. D. Thompson, Phys. Rev. Lett. **89**, 106402 (2002)

[8] E. J. Singley, D. N. Basov, E. D. Bauer, and M. B. Maple, Phys. Rev. B **65**, 161101(R) (2002)

[9] R. Settai, H. Shishido, S. Ikeda, Y. Murakawa, M. Nakashima, D. Aoki, Y. Haga, H. Harima, and Y. Onuki, J. Phys.: Condens. Matter **13**, L627 (2001)

[10] V. A. Sidorov, M. Nicklas, P. G. Pagliuso, J. L. Sarrao, Y. Bang, A. V. Balatsky, and J. D. Thompson, Phys. Rev. Lett. **89**, 157004 (2002)

[11] Johnpierre Paglione, M. A. Tanatar, D. G. Hawthorn, Etienne Boaknin, R. W. Hill, F. Ronning, M. Sutherland, Louis Taillefer, C. Petrovic, and P. C. Canfield, Phys. Rev. Lett. **91**, 246405 (2003)

[12] Donavan Hall, E. C. Palm, T. P. Murphy, S. W. Tozer, Z. Fisk, U. Alver, R. G. Goodrich, J. L. Sarrao, P. G. Pagliuso, and Takao Ebihara, Phys. Rev. B **64**, 212508 (2001)

[13] P. M. Rourke, M. A. Tanatar, C. S. Turel, J. Berdeklis, C. Petrovic, and J. Y. Wei, Phys. Rev. Lett. **94**, 107005 (2005)

[14] M. A. Tanatar, Johnpierre Paglione, S. Nakatsuji, D. G. Hawthorn, E. Boaknin, R. W. Hill, F. Ronning, M. Sutherland, Louis Taillefer, C. Petrovic, P. C. Canfield, and Z. Fisk, Phys. Rev. Lett. **95**, 067002 (2005)

[15] E. D. Bauer, F. Ronning, C. Capan, M. J. Graf, D. Vandervelde, H. Q. Yuan, M. B. Salamon, D. J. Mixson, N. O. Moreno, S. R. Brown, J. D. Thompson, R. Movshovich, M. F. Hundley, J. L. Sarrao, P. G. Pagliuso, and S. M. Kauzlarich, Phys. Rev. B 73, 245109 (2006)

[16] G. Zwicknagl, Adv. Phys. **41**, 203 (1992)

[17] G. Goll, J. Hagel, H. V. von Lohneysen, T. Pietrus, S. Wanka, J. Wosnitza, G. Zwicknagl, T. Yoshino, T. Takabatake, A. G. M. Jansen, Europhys. Lett. **57**, 233 (2002)

[18] N. Kozlova, J. Hagel, M. Doerr, J. Wosnitza, D. Eckert, K.-H. Müller, L. Schultz, I. Opahle, S. Elgazzar, Manuel Richter, G. Goll, H. v. Löhneysen, G. Zwicknagl, T. Yoshino, and T. Takabatake, Phys. Rev. Lett. **95**, 086403 (2005)

[19] S. Elgazzar, I. Opahle, R. Hayn, and P. M. Oppeneer, Phys. Rev. B **69**, 214510 (2004)

[20] N. Harrison, U. Alver, R. G. Goodrich, I. Vekhter, J. L. Sarrao, P. G. Pagliuso, N. O. Moreno, L. Balicas, Z. Fisk, D. Hall, Robin T. Macaluso, and Julia Y. Chan, Phys. Rev. Lett. **93**, 186405 (2004)

[21] Yoshinori Haga, Yoshihiko Inada, Hisatomo Harima, Kenichi Oikawa, Masao Murakawa, Hirokazu Nakawaki, Yoshihumi Tokiwa, Dai Aoki, Hiroaki Shishido, Shuugo Ikeda, Narumi Watanabe, and Yoshichika Ōnuki, Phys. Rev. B **63**, 060503 (2001)

[22] Donavan Hall, E. C. Palm, T. P. Murphy, S. W. Tozer, C. Petrovic, Eliza Miller-Ricci, Lydia Peabody, Charis Quay Li, U. Alver, R. G. Goodrich, J. L. Sarrao, P. G. Pagliuso, J. M. Wills, and Z. Fisk, Phys. Rev. B **64**, 064506 (2001)

[23] D. P. Moore et al., Physica B **312-313**, 134 (2002)

[24] Shin-ichi Fujimori, Tetsuo Okane, Jun Okamoto, Kazutoshi Mamiya, Yasuji Muramatsu, Atsushi Fujimori, Hisatomo Harima, Dai Aoki, Shugo Ikeda, Hiroaki Shishido, Yoshifumi Tokiwa, Yoshinori Haga, and Yoshichika Ōnuki, Phys. Rev. B **67**, 144507 (2003)

[25] Shin-ichi Fujimori, Atsushi Fujimori, Kenya Shimada, Takamasa Narimura, Kenichi Kobayashi, Hirofumi Namatame, Masaki Taniguchi, Hisatomo Harima, Hiroaki Shishido, Shugo Ikeda, Dai Aoki, Yoshifumi Tokiwa, Yoshinori Haga, and Yoshichika Ōnuki, Phys. Rev. B **73**, 224517 (2006)

[26] S. Raj, Y. Iida, S. Souma, T. Sato, T. Takahashi, H. Ding, S. Ohara, T. Hayakawa, G. F. Chen, I. Sakamoto, and H. Harima, Phys. Rev. B **71**, 224516 (2005)

[27] A. Koitzsch, S. V. Borisenko, D. Inosov, J. Geck, V. B. Zabolotnyy, H. Shiozawa, M. Knupfer, J. Fink, B. Büchner, E. D. Bauer, J. L. Sarrao, and R. Follath, Phys. Rev. B **77**, 155128 (2008)

[28] K. Koepernik and H. Eschrig, Phys. Rev. B **59**, 1743 (1999)

[29] http://www.fplo.de

[30] S. Hüfner, *Springer Series in Solid State Sciences Bd. 82: Photoelectron Spectroscopy*, Berlin, Springer Verlag (1995)

[31] J. J. Yeh and I. Lindau, Atom. Data Nucl. Data Tabl. **32**, 1 (1985)

[32] J. D. Denlinger, G. H. Gweon, J. W. Allen, C. G. Olson, M. B. Maple, J. L. Sarrao , P. E. Armstrong, Z. Fisk, and H. Yamagami, J. Electron Spectr. Rel. Phen. **117-118**, 347 (2001) Rev. Lett. **97**, 117002 (2006)

[33] Satoru Nakatsuji, David Pines, and Zachary Fisk, Phys. Rev. Lett. **92**, 016401 (2004)

# Captions:

**Fig. 1**: Fermi surface map at T = 25 K with hν = 100 eV photon energy. Bright yellow corresponds to high photoemission intensity. The first Brillouin zone is highlighted by dashed lines. The Fermi surface consist of a "flower" structure around the zone center and closed "square"-like structures around the corners. Theoretical FS are shown in the upper corner for the Z-plane (green) and for the Γ-plane (blue). The f electrons have been treated as localized. Closer inspection of the square reveals two bands α and β (see Fig. 5). An additional "line" feature is found near the second Z-point ($Z_2$).

**Fig. 2**: a) Experimental energy distribution map along Z-A (see Fig. 1). b) – d) Co-3d orbital projected LDA calculations for BZ cuts along Z-A and Γ-M where the 4f electrons have been treated as fully localized or itinerant. "F" denotes the flower structure. e) Intermediate cut parallel to Z-A with $k_z = 0.85\ \pi/c$.

**Fig. 3**: a) Experimental energy distribution map along Z-R (see Fig. 1). b) – d) Co-3d orbital projected LDA calculations for BZ cuts along Z-R and Γ-X where the 4f electrons have been treated as fully localized or itinerant. "F" denotes the flower structure.

**Fig. 4**: a) Experimental energy distribution map along A-R (see Fig. 1). b) – d) Co-3d orbital projected LDA calculations for BZ cuts along A-R and M-X where the 4f electrons have been treated as fully localized or itinerant. The red arrow highlights the position of the "line" feature. (e) Experimental energy distribution map taken with $h\nu = 87$ eV.

**Fig 5**: Series of EDMs 1-6 as denoted in Fig.1. The splitting in α and β sheet becomes apparent as one moves away from the high symmetry direction. The distance between the bands increases. An additional "line" feature is observed at lower k which corresponds to the accordingly labeled sheet in Fig. 1.

Fig. 1

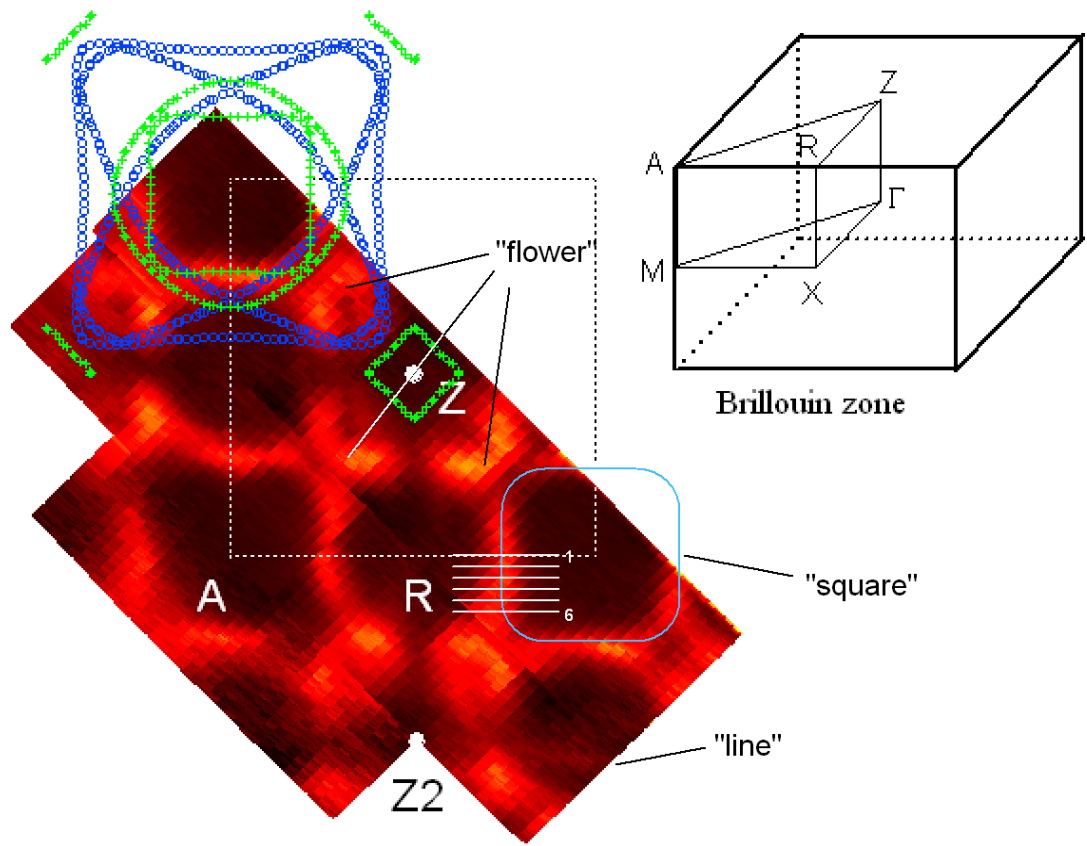

Fig. 2

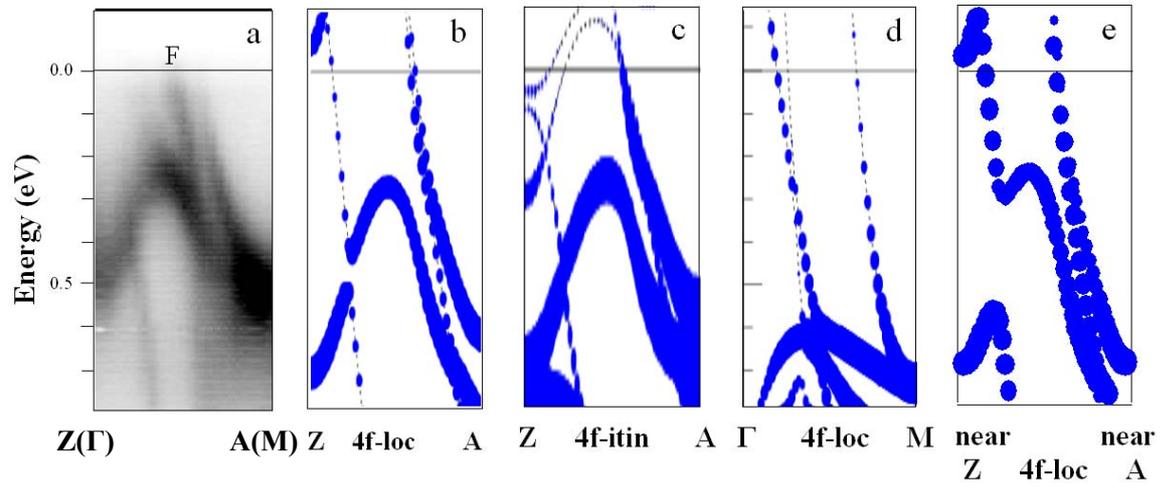

Fig. 3

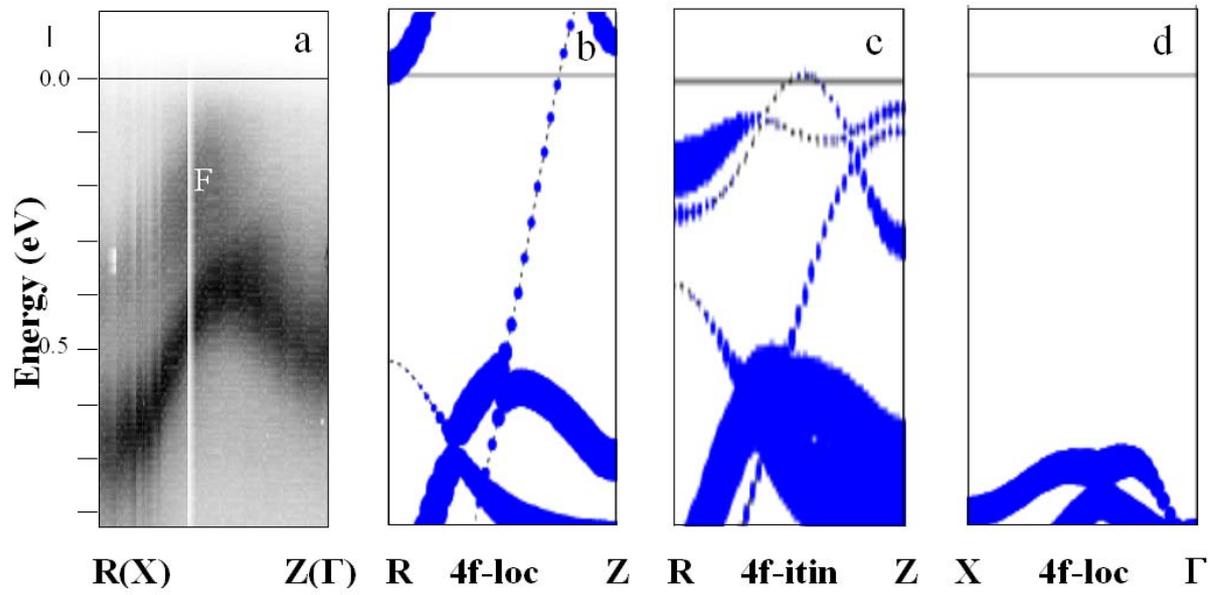

Fig. 4

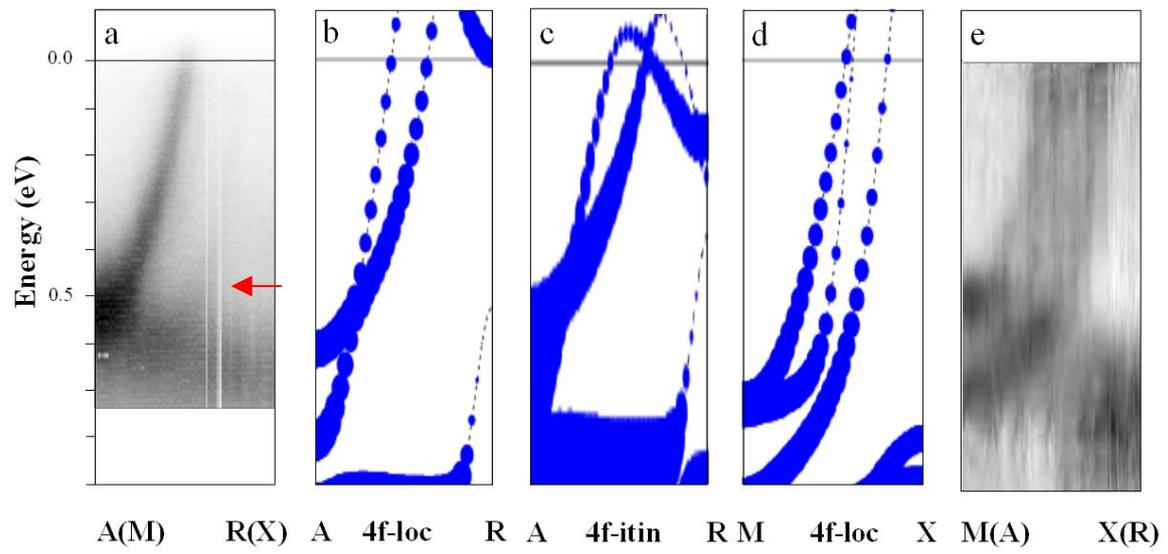

Fig. 5

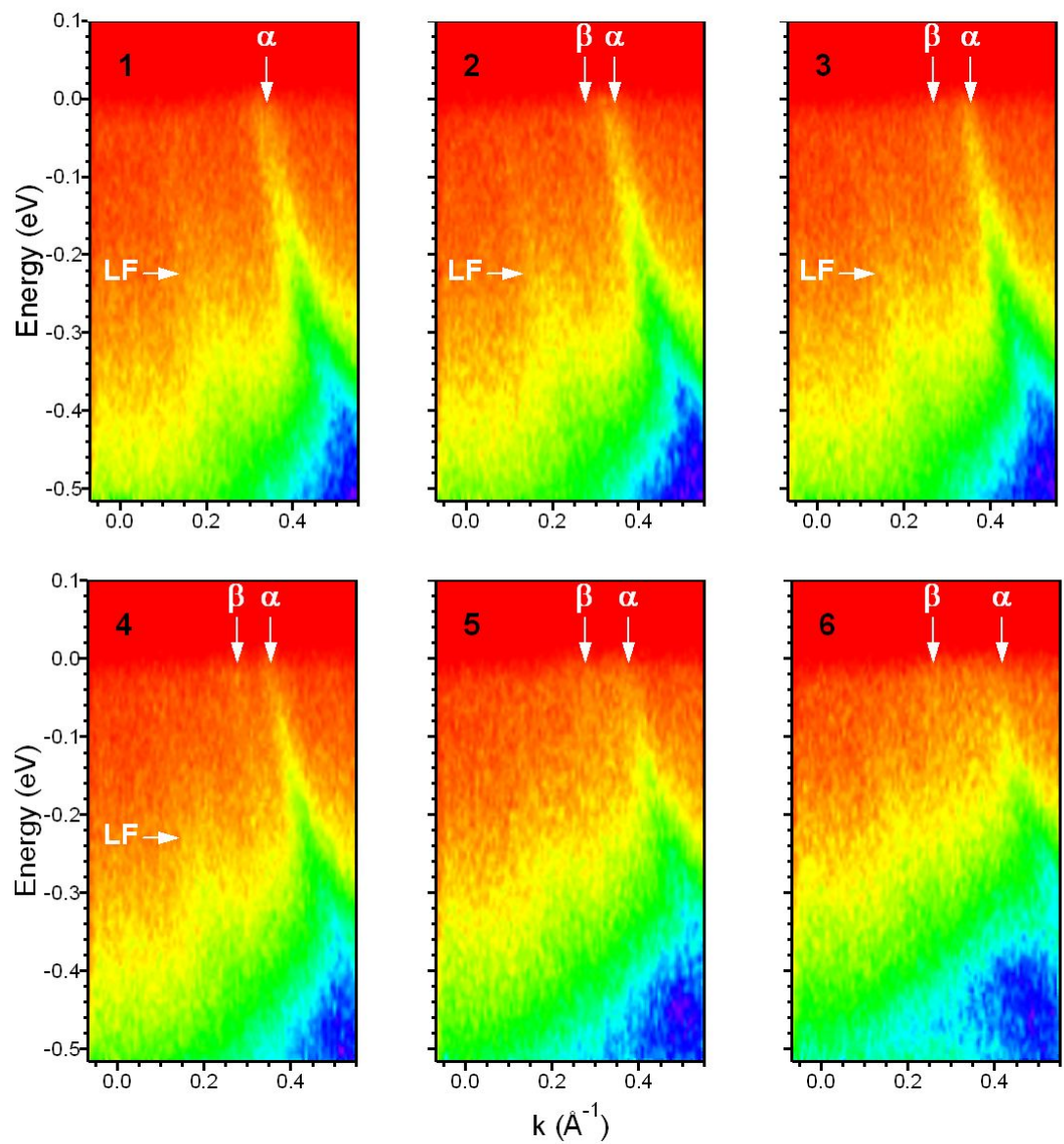